# Electric field distortions in structures of the twist bend nematic ($N_{TB}$) phase of a bent-core liquid crystal

K. Merkel*[1], A. Kocot[1], J. K. Vij[§2], and G. Shanker[3]

[1]Institute of Material Science, Silesian University, Katowice, Poland
[2]Department of Electronic and Electrical Engineering, Trinity College Dublin, the University of Dublin, Dublin 2, Ireland
[3]Department of Studies in Chemistry, Bangalore University, Bangalore, India

**ABSTRACT**

Dielectric spectroscopy of a twist–bend nematic phase ($N_{TB}$) of an achiral bent–core liquid crystalline compound under DC bias is used to investigate its response to electric field. Two collective relaxation processes are revealed, these are assigned to distortions of helicoidal structure by the external bias field. The mode at a frequency centred at ~ $10^5$ Hz is assigned to the local distortion of the helicoidal angle while the periodic helical structure of $N_{TB}$ remains unaltered at this time-scale. Frequency of the mode depends primarily on the helicoidal angle and has anomalous, softening- like behaviour at the N-$N_{TB}$ transition. The second process is observed in the frequency range below 10 kHz. A coupling of dielectric anisotropy with electric field gives rise to a new equilibrium periodic structure in the time scale involved. The modulus of the wave vector gradually vanishes on increasing the bias field (except for the initial behaviour, $E^2 < 0.1$ $V^2/\mu m^2$, which is just the opposite). Transition from the twist–bend to the splay-bend structure is clearly observed by a sudden drop in the frequency of this mode, which decreases almost linearly with increasing field. Results agree with predictions from current models for the periodically distorted $N_{TB}$ phase.

*merkelkatarzyna@gmail.com

Two remarkable physical phenomena for bent-core liquid crystals have recently been discovered. These arise from the polar order and bending of the director. In addition to these four nematic phases already known to exist: (i) conventional, (ii) biaxial, (iii) blue, and (iv) cholesteric, the twist-bend nematic ($N_{TB}$) as the fifth in sequence has recently been discovered [1-4]. In this phase the director has a uniform bend and a uniform twist and the optical axis coincides with the helical axis. The helicoidal angle is less than $\pi/2$. This phase is triggered by-unusual elastic properties of strongly bent shaped molecules. If the polar order exists along the orthogonal direction then net polarization emerges and it interacts with electric field. Meyer and Dozov [5] recently showed that the polar order in the transverse directions couples to the bend variations of the main director. Response of the $N_{TB}$ to the external stimuli such as the applied electric field or chiral doping [6,7] can yield information about detailed structure of the phase. The phenomenon dealing with formation of $N_{TB}$ has the potential for practical applications, once the structure of the phase and the reasons for its formation are well understood. We can then design new materials to tailor to the specific requirements. Electric field, as the external stimulus, has important advantages as this couples directly to dielectric anisotropy as well to the polar order. Furthermore electric field can conveniently be applied across a liquid crystal cell. These effects arise from distortions in the polar order induced by the field and the interaction of the field with the spontaneous polarization. The latter are governed by flexoelectricity and ferroelectricity, respectively.

The question as to whether the nematic phase displayed by achiral bent-core systems displays ferroelectric effects is currently being debated [8,9] but this discussion is not yet extended to $N_{TB}$ phase. In this the effect of electric field, $E$, can be observed through strong dielectric and weak flexoelectric couplings [6,7]. A linear electro-optic effect is observed in $N_{TB}$ where the optical axis is rotated by $E$ applied in a plane perpendicular to the helical axis. Such a characteristic effect resembles the electro-clinic effect in SmA* [10] and flexoelectric effect (FEE) in cholesteric nematic (N*) [11]. For the case when dielectric anisotropy, $\Delta\varepsilon < 0$, in the $\mathbf{E} \perp \mathbf{q}$ geometry of an electroclinic experiment on a planar-aligned sample, the average dielectric torque acting on the helical axis is zero. The flexoelectric effect though independent of dielectric one is operative here. When the high frequency AC field is applied across a planar-aligned cell, the dielectric effect due to $\Delta\varepsilon$ is dominant even for weak fields, due to finite time average of the torque that acts on the helical axis. We can neglect a variation in the wavevector $\boldsymbol{q} = 2\pi/\boldsymbol{p}$ by $E$, by assuming that pitch $\boldsymbol{p}$ is dynamically frozen at its field-off value [12]. At lower frequencies, however, we have to account for the macroscopic effects that arise from *E*. These are related to the reorganisation of the pseudo-layers in the structure of the $N_{TB}$ phase and to nucleation and propagation of defects. Results are analyzed in terms of models given by Matsuyama [13] and Pająk et al. [14-15].

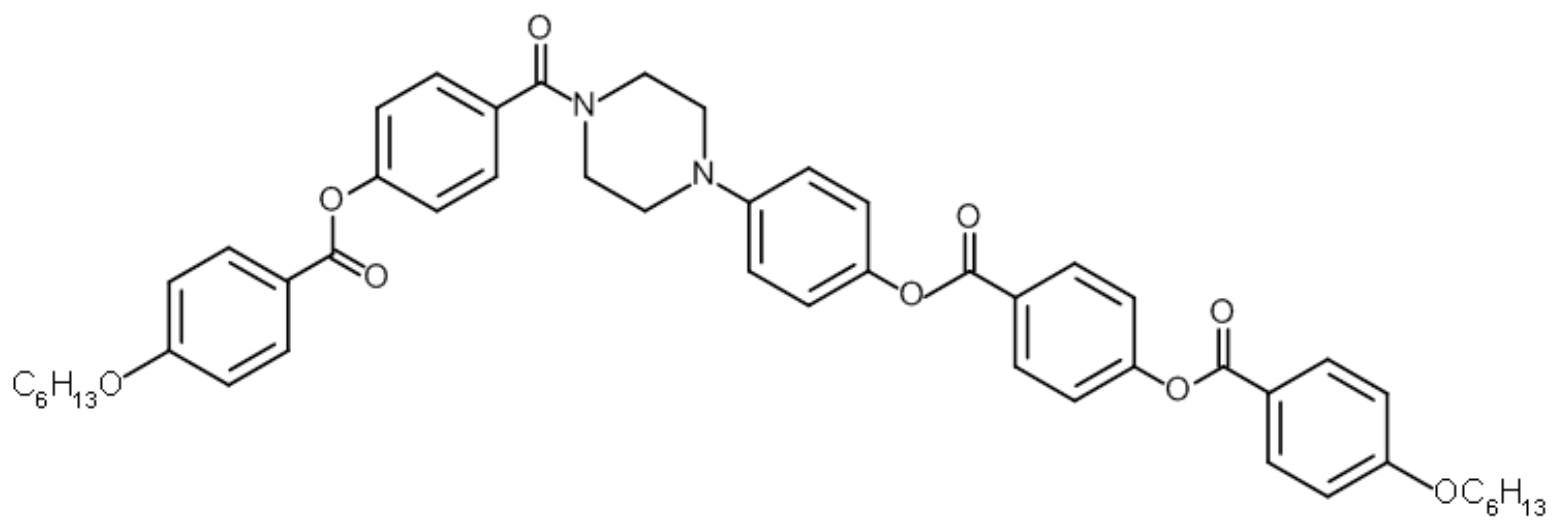

**Cr-** 392 K- **Col** -423 K- $\mathbf{N_{tb}}$ 439 K- **N** -462 K - **Iso**

Fig.1 Chemical structure and transition temperatures of the bent core molecule utilized for the present study.

In this Letter, results of experiments on dielectric spectroscopy with and without bias field and of electro-optics are given and the data analysed. Dielectric spectra are expressed in terms of the relaxation processes: both molecular and collective. The latter are related to (a) the tilt or the helicoidal angle fluctuations, and (b) slow distortions of $N_{TB}$ structure induced by external field. Results are analysed in terms of current models for the $N_{TB}$ phase subject to the electric field. The dielectric spectra ε" of a planar aligned LC cell are shown in Figs. 2 and 3a. The real and imaginary parts of the complex permittivity were measured in the frequency range of 1 Hz to 100 MHz, in planar and homeotropic cells under slow cooling, with cell-spacing ranging from 1.6 μm to 10 μm.. For high frequency measurement, the samples were filled in golden plated cells (AWAT, Poland) with proper alignment layers. Amplitude of the measuring field lays the range 0.01- 1V/μm, and DC bias field up to 8 V/μm was applied across the cell. The spectra reveal two relaxation peaks in the high frequency range, shown in Fig. 2a; related to the two molecular relaxation processes. The two additional low frequency relaxation processes shown in Figs. 2b and 3a are assigned to the collective processes. Similarly the two peaks observed at higher frequencies in the dielectric spectra are recorded from a homeotropic-aligned cell.

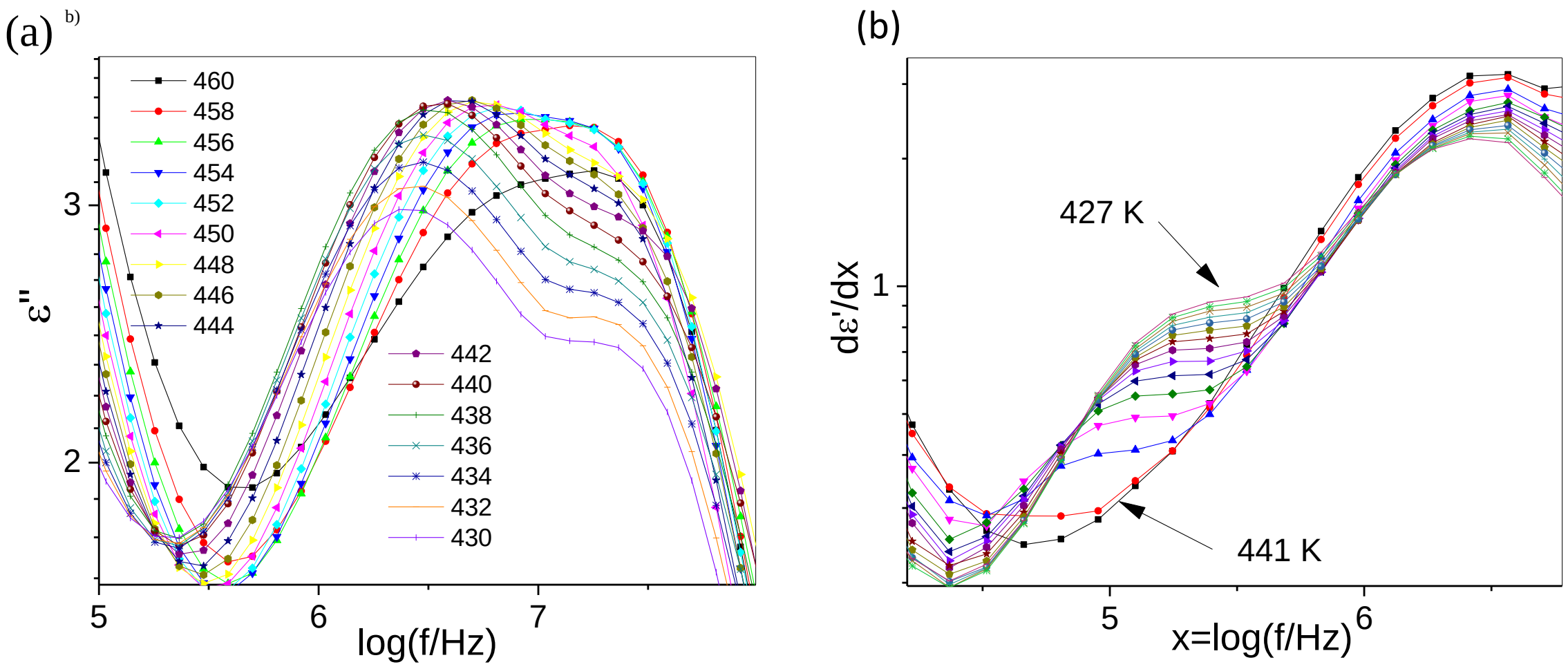

Fig. 2. The dielectric spectra recorded for a planar aligned 5 μm sample. (a) $\varepsilon''$ in the frequency range of molecular modes (b) derivative: d$\varepsilon'$/d log($f$) plotted as a function of frequency in the low frequency range.

In order to describe the various dielectric modes, dielectric strength and relaxation time of each process is calculated using Cole-Cole equation that expresses complex dielectric permittivity in terms of the relaxation times of various processes, their dielectric strengths and symmetrical distribution parameters.

$$\varepsilon^*(\omega) - \varepsilon_\infty = \sum_{j=1}^{n} \frac{\delta\varepsilon_j}{[1+(i\omega\tau_j)^\alpha]}, \qquad (1)$$

Results of fitting reveal that dielectric spectra displayed by achiral bent-core molecules are much more complicated than for systems composed of rigid core molecules in the nematic phase. Thus for achieving a better deconvolution of relaxation peaks in frequency range of $10^4$ - $10^7$ Hz, it is preferable to analyze the derivative of the real part of permittivity ε′ [16] as follows:

$$\frac{d\varepsilon'}{d(\ln f)} = \frac{d\varepsilon'}{d(\ln\omega)} = \sum_{j=1}^{n} \mathrm{Re}\frac{\delta\varepsilon_j \alpha (i\omega\tau_j)^\alpha}{[1+(i\omega\tau_j)^\alpha]^2}, \qquad (2)$$

Rates of the various relaxation processes in terms of frequencies are plotted vs temperature in Fig. 3b. Frequency dependent dielectric permittivity results for planar and homeotropic aligned cells are analyzed in terms of the Maier and Meier (M-M) model as modoified by Toriyama et al. [17].

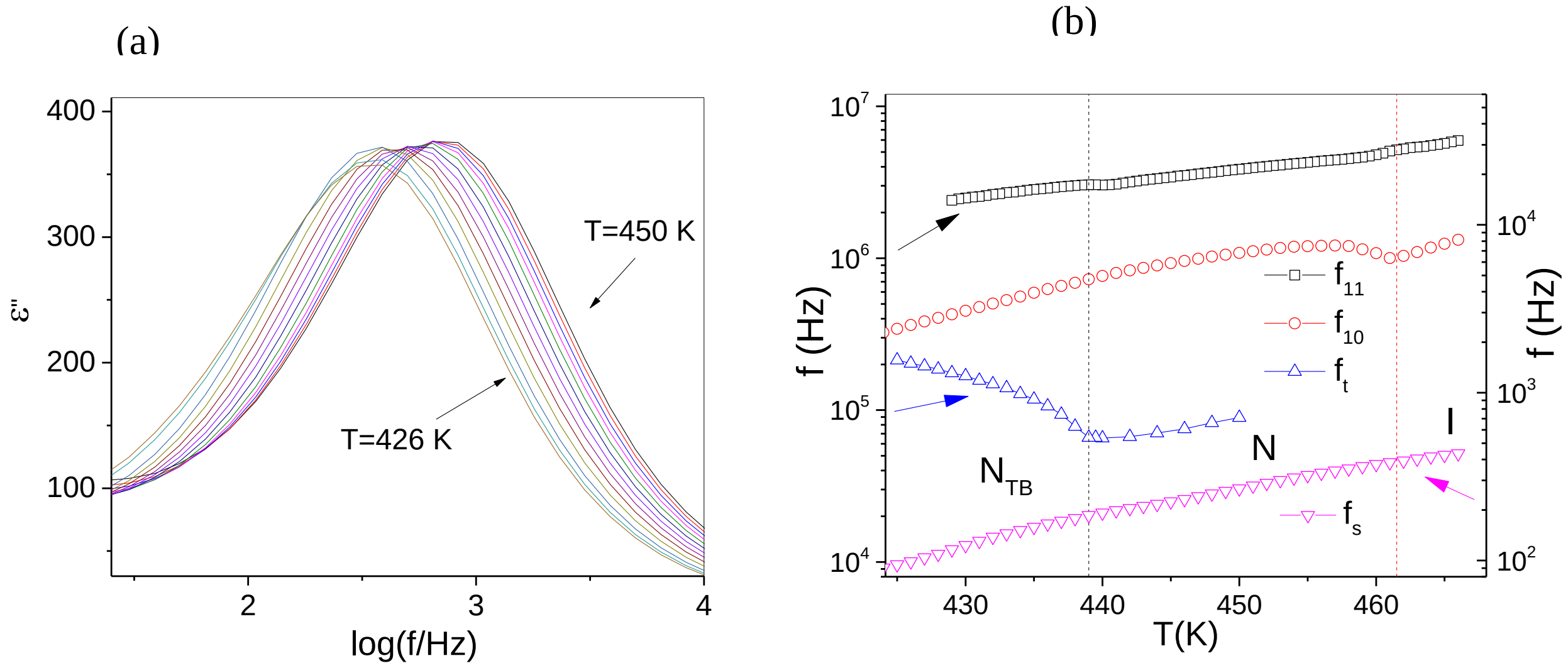

Fig. 3a (color online) The dielectric loss spectra for a planar aligned cell of cell-spacing 5 μm, with a temperature step of 2 K are plotted as a function of the logarithmic frequency and for different temperatures (b) Relaxation frequencies corresponding to the molecular modes: precessional, $f_{10}$, □, spinning, $f_{11}$, O, and the $N_{TB}$ tilt deformation mode, $f_t$, Δ, the low frequency mode $f_S$ obtained from a planar-aligned cell of cell spacing 5μm,∇.

Analysis of the frequency dependence of the dielectric permittivity given for the rotational diffusion model [18] is also attempted and compared with the experiment.

$$\varepsilon_{\perp}^{*}(\omega)-n_{\perp}{}^{2}\approx\frac{N'hF^{2}g_{\perp}}{3\varepsilon_{0}k_{B}T}\frac{1}{3\mu_{0}^{2}}\left[\frac{\mu_{l}{}^{2}(1-S)}{1+i\omega\tau_{10}}+\frac{\mu_{t}{}^{2}(1+\frac{1}{2}S)}{1+i\omega\tau_{11}}\right] \qquad (3).$$

In eqn. (5), $\mu_l$ and $\mu_t$ are the projections of the longitudinal and the transverse components of the molecular dipole moment directed along and normal to the long molecular axis, $\mu_0$ is the total dipole moment of the molecule, $\tau_{10}$ and $\tau_{11}$ are the relaxation times that describe the precessional and the spinning rotations of the LC molecules, respectively. Expressions for the ratios $\frac{\tau_{10}}{\tau_0}$ and $\frac{\tau_{11}}{\tau_0}$ are described in terms of the orientational order parameter $S$ and the anisotropy in diffusion coefficients Δ.

$$\frac{\tau_{10}}{\tau_{0}}=\frac{1-S}{1+\frac{1}{2}S},\quad \frac{\tau_{11}}{\tau_{0}}=\frac{2+S}{2+\Delta(1-\frac{1}{2}S)}, \qquad (4)$$

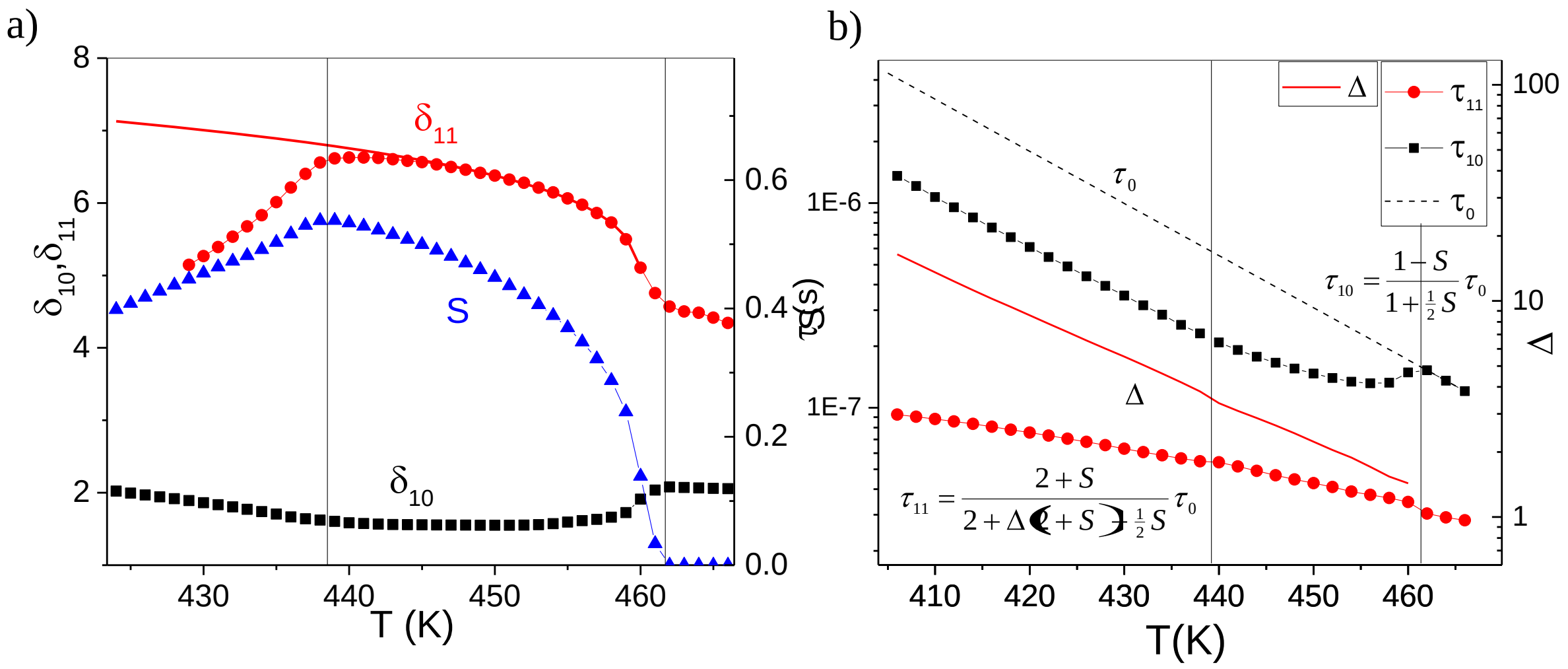

Fig. 4. (a) Plots of dielectric strengths of molecular modes: precessional, $\delta_{10}$, ■, spinning , $\delta_{11}$, ●. $\delta_{11}$ corresponding to the Mayer-Saupe model extrapolated down to low temperatures. The orientational order parameter, S, ▲, (b) relaxation time of precessional, $\tau_{10,}$ ■, and spinning, $\tau_{11}$, ● modes. The extrapolation of the relaxation time from the isotropic phase as a dotted line and anisotropy in the rotational diffusion coefficients marked as a red continuous line, Δ, in Fig. 4b are shown as a function of temperature.

For the set of equations (4) given above, $\tau_0$ is the relaxation time of the system in its isotropic phase. $\tau_0 = \zeta / k_B T$, where, $\zeta$, is the coefficient of friction and $\Delta = \frac{1}{2}[D_{\parallel} / D_{\perp} - 1]$ is the anisotropy in the rotational diffusion coefficients of $D_{\parallel}$ and $D_{\perp}$, as defined above. Figures 4a and 4b show dependencies of dielectric strengths and relaxation times on temperature for the two processes that contribute to the dielectric spectra at higher frequencies ($10^6$ Hz - $10^8$ Hz). For the conventional nematic phase, we find that results are well reproduced by rotational diffusion model [18]. Temperature dependency of orientational order parameter, *S*, obtained from results of dielectric relaxation strength are plotted in Fig. 4a. An anisotropy in the rotational diffusion, *Δ*, plotted in Fig. 4b, was calculated from the relaxation time. Temperature dependence of *S* distinctly reverses its monotonic trend of increase to decrease at the transition temperature of 438 K, but the curve nevertheless departs from the expected Maier-Saupe behavior at a temperature few degrees above N to $N_{TB}$ transition (Fig. 4a). This shows that the director is tilted on the cone not only at a temperature of 438 K but the tilt emerges a few degrees above the transition temperature in the high temperature nematic phase. The rotational viscosity, $\gamma$, for rotation of local nematic director is calculated from $\tau_{0,}$ by using the relation:

$$\tau_0 = \frac{\zeta}{kT} = \frac{3\gamma V}{kT} \qquad (5).$$

In Eqn. (5) $V$ is the volume of a single LC molecule. $\tau_0$ is the experimentally measured relaxation time in the isotropic phase and is related to the rotational viscosity $\gamma$ through Eq. (5). On using eqn. (4) for ratio of $\frac{\tau_{10}}{\tau_0}$ as found here, we can also calculate $S$ here too. For the general case of the electric field being sufficiently large, dielectric coupling dominates expression of the free energy density of the twist-bend nematic phase. In addition, when the high frequency AC signal is applied, the dielectric effect is even more dominant for weak fields due to the finite time average of the torque for the quadratic effect that acts on the **q** axis, in contrast to the polar effect for which the time average of torque is zero. In most cases however, the dielectric effect reorients the macroscopic symmetry axis, ***N***, as a result of the sum of the torques acting locally on the director ***n***.

The reorganisation of pseudo-layered structures of the twist-bend nematic is slow and in order to avoid complications of the LC dynamics, we assume that $E$ varies rapidly on the millisecond time scale while discussing the high frequency mode. In that case, we can safely neglect variations of ***q*** with time by assuming that helical pitch is dynamically frozen in its field-off value. We consider the particular case when field stabilises the macroscopic heliconical structure [12].

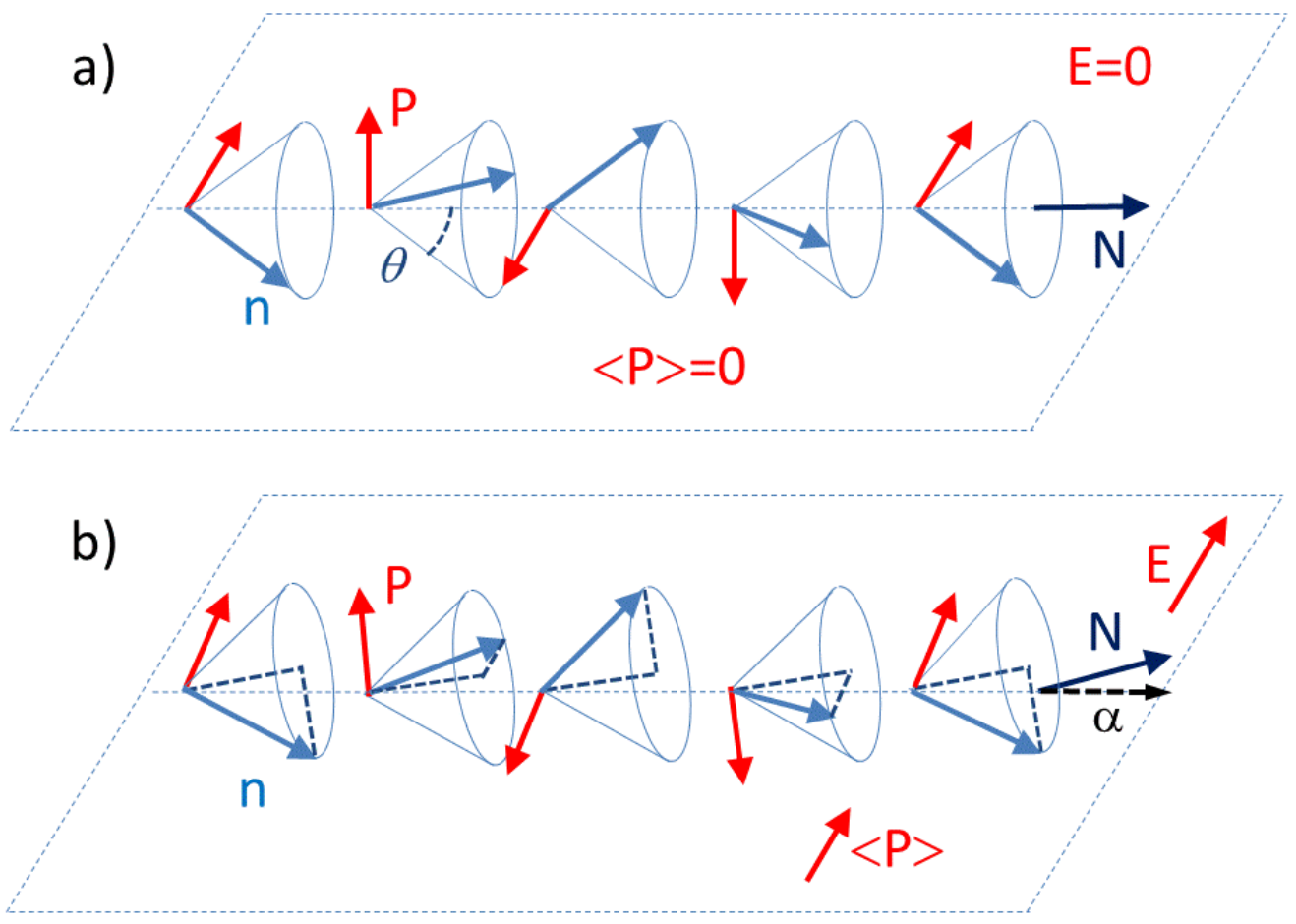

Fig.5. Demonstration of the electroclinic effect in $N_{TB}$: (a) without field, **n** is tilted and twisted and follows heliconical structure, (b) Optical or the symmetry axis **N** rotates around E with an angle α.

The field-induced distortional effect in $N_{TB}$ may be extremely useful for technological applications but may be similar to predictions and observations of the electroclinic effect [19,20].

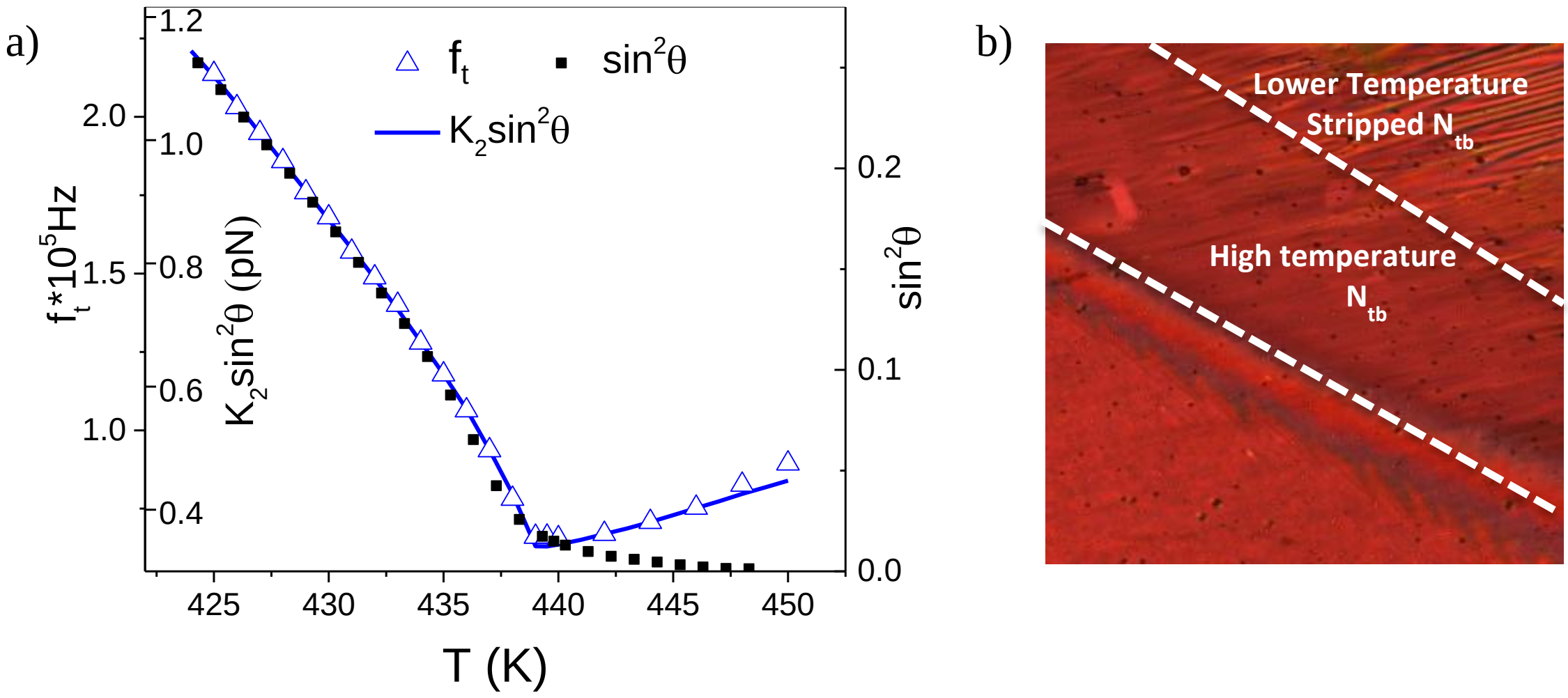

Fig.6. (a) Relaxation frequency of the helix tilt deformation mode, $f_t$ , Δ. Continuous line is the product of twist elastic constant $K_2$ and $\sin^2\theta$. $K_2*\sin^2\theta$ is obtained as an unknown parameter by fitting of eq. (9) to the experimental data of $f_t$. $\sin^2\theta$ is found as an angle of rotation of the dielectric tensor, ■. $\theta$ is the helicoidal angle. (b) The optical polarizing microscopic textures for N and $N_{TB}$ phase.

Figure 6a shows dependence of the relaxation rate of the tilt distortion mode, $f_t$, plotted as a function of temperature for $N_{TB}$. This dependence observed from dielectric spectra appears similar to that of director fluctuations from dynamic light scattering by Parsouzi et al [21] of a thermotropic liquid crystalline mixture of a monomer mixed with a dimer. The relaxation time in the $N_{TB}$ phase is found to vary from 2 to 0.7 μs from 438K to 425K. This is obtained from a fitting of the spectra, shown in Fig 2b, to Eqn. (1). Values so obtained are found to be much lower than those obtained from optical switching (15 μs to 11 μs) [20]. On the contrary, relaxation time is rather similar to the optical switching time, ~0.7 μs for CB7CB [4,12] and C11CB [19]. We try to reproduce temperature dependencies of the high frequency mode centred at ~ $10^5$ Hz using tilt's response time [12]:

$$\tau_t = \frac{\gamma}{4q^2K_2\sin^2\theta} \qquad (6)$$

$\gamma$ is the rotational viscosity of the local nematic director $\boldsymbol{n}$ in the nematic phase, and is expected to be continuous at the N–$N_{TB}$ transition. Qualitatively, $\gamma$ and $q^2$ have similar temperature dependencies and hence $\gamma /q^2$ is a constant in eqn. (6). Pitch pf the $N_{TB}$ is assumed as $p_0 \approx 14$ nm [22]. $\gamma$ is calculated from $\tau_{10}$, of the precessional mode is found as ~ 0.9 Pa·s. The conical angle is calculated from temperature dependencies of the dielectric data on the assumption that $S$ in the uniform nematic phase follows the Meier-Saupe theory [17]. The conical angle reaches 30° on approaching 425 K from above. This is much greater than

~$10^o$ found from birefringence measurements [20] but is similar to that for CB7CB dimer [23]. Figure 6a shows that in $N_{TB}$ phase, the product, $K_2 \sin^2\theta$, obtained from the fitting of eq. (6) can also be reproduced by $\sin^2\theta$, $\theta$ here is found from dielectric data. Both curves have quite similar temperature dependencies but their departures from each other above the $N_{TB}$-N transition temperature can be explained by special temperature dependence of the bend elastic constant close to the transition temperature, which may be similar to CB7CB [24]. It is interesting to note that conical angle is already finite at few degrees above the $N_{TB}$-N transition temperature. This temperature range coincides with the so-called high temperature $N_{TB}$ phase [20], as shown in Fig. 6b.

The lowest frequency relaxation process is observed in the range ($10^2$ Hz - $10^4$ Hz). This is related to the spontaneously formed $N_{TB}$ structure under the influence of $E$.–The process can be generated by the dielectric and/or flexoelectric effects and appears to be related to transverse domains of submicron periodicity. The process can be assigned as a hydrodynamic twist-bend director mode with relaxation rate that is expected to be $q^2$ dependent. where $\boldsymbol{q}=q\boldsymbol{z}$ , with z-dependent rotation of $\boldsymbol{n}(\boldsymbol{r})$ and z-dependent displacement of pseudolayers (leading to compression and dilation) [21]. The above process, Figs 3a and 3b, is characterized by a large dielectric strength, $\delta\varepsilon \cong 900$, indicative of a significantly large polar order. Frequency, $f_S$, of the mode is strongly influenced by surface interactions and external electric field. Both of these stimuli increase the period and reduce the relaxation frequency. It is possible to predict periodicity, $p$, or wave vector, $q$, of the transverse pattern using expression for $f_S$ in the single elastic constant approximation:

$$f_s = \frac{q^2 K}{\gamma_s} \qquad (7).$$

The viscosity coefficient, $\gamma_s$, is related to the reorientation of the polar axis, i.e. rotation of the perpendicular component, $\mu_t$, due to spinning. Thus, $\gamma_s$, is described by $\tau_{11}$ and can be calculated using eqn (4). The predicted domain periodicity, found for a given elastic constant ($K \approx 4$pN) lies in the submicron scale (200-900 nm). Such sub-micrometer patterns were already observed using three-photon excitation fluorescence polarizing microscopy of hydrocarbon-linked mesogenic dimers [25,26].

Further important characteristic of a modulated nematic structure is the variation of the wave vector with field. Such an approach is presented recently by Pająk *et. al* [14], as response of the bulk $N_{TB}$ phase to external electric fields within the framework of the Landau-de Gennes free energy expression. As $N_{TB}$ is expected to occur in nonchiral bent–shaped molecules, with and without electric dipoles, Pajak et al assume that stability of this phase is driven primarily by the excluded–volume entropic interactions. They consider Landau-de Gennes type free energy expansion, $F_E$ , [27] in terms of the traceless tensor order parameter $Q$(r) and polarization field $P$(r) produced by the electric field.

$$F_E = -\Delta\varepsilon \frac{1}{V} \int E_\alpha Q_{\alpha\beta}(r) E_\beta d^3 r \qquad (8)$$

$\Delta\varepsilon$ is the dielectric anisotropy in the director reference frame. They assumed that the dielectric term dominates, at least for sufficiently large fields. The relative orientations of $\boldsymbol{Q}$ and field $\boldsymbol{E}$ are parameterized by $\theta$ and $\varphi$. These are found by minimization of $F$ with respect to $\theta$ and $\varphi$. Here $F = F_N + F_E$, where $F_N$ is the free energy of the various nematic phases. This entropically induced state is realized since the original molecules have a specific shape. For samples with negative dielectric anisotropy, the transformation of the initial phase is possible in three types of one-dimensional modulated nematic structure (ODMNS) with a wide field range of the $N_{SB}$ phase. This phase is stabilized by the application of $E$. We can make use of the general concept that external field distorts $N_{TB}$ structure and $f_S$ given by eqn. (7) is proportional to $q^2$. Then we can compare the experimental dependence of the $q^2_{exp}$ vs. $E$ from the dielectric data and from the texture with $q^2$ vs. $E$ predicted by the model [14]. Figure 7a show the experimental dependence of $p_{exp}{}^2$ with respect to $E$. Clearly, the wave vector gradually vanishes on increasing the field and the frequency of the mode decreases almost linearly with the electric field. The general effect of the field is to unwind the structure and to reduce frequency of the mode, except for an initial behaviour ($E^2 < 0.1$ V$^2$/μm$^2$), which is just the opposite to that subsequently observed. Such behaviour at low fields is reminiscent of the Goldstone mode in FLCs. The case of negative dielectric anisotropy modifies modulation of the main director, which is then undergoing precession on the elliptic cone around $\boldsymbol{q}$, the elliptic twist bend, $N_{TBe}$, phase is displayed. Stronger fields in the ($\boldsymbol{E}\perp\boldsymbol{q}$) geometry make the base of the elliptic cone narrower and finally $N_{SB}$ is stabilized. The cone is degenerated to a line with $P$ in the plane of splay–bend modulations. Larger field diminishes modulations in–plane with polarization $P$ obtaining the off-plane uniform component along the field. This new chiral structure is denoted by $N^*_{SBp}$. For even stronger fields, this phase transforms to polar nematic, $N_p$ with polarization directed parallel to the field.

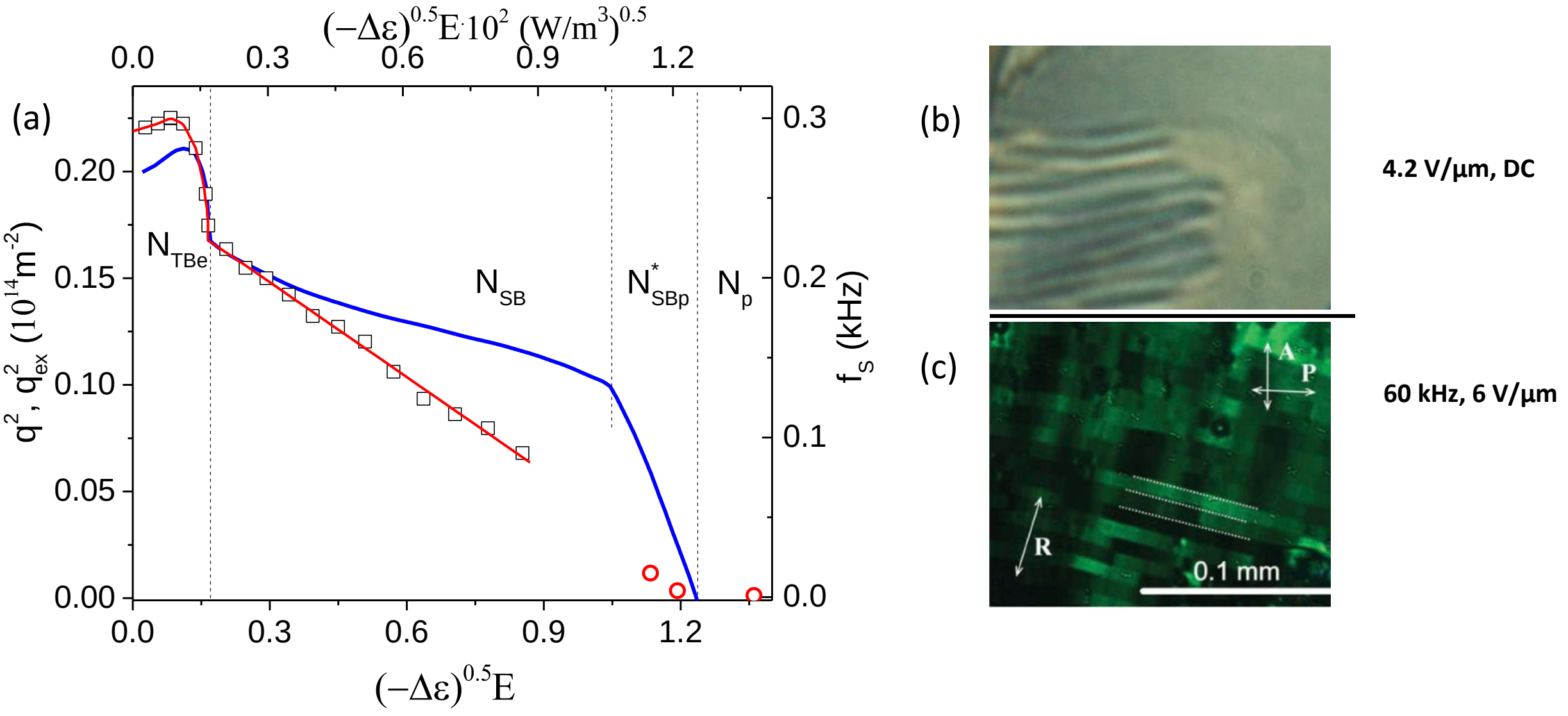

Fig.7. (a) Wave vector, $q^2$, plotted as a function of $E$ for ODMNS as the outcome of free energy minimization [14] shown as as a blue line. (X axis bottom, Y axis, the left scale). Peak frequencies and wave vector for the low frequency mode at T=425 K, dielectric data, $f_S$, □, and from texture, ○, as a function of the bias field E (X axis top, Y axis, right hand scale). (b) Texture obtained by application of DC electric field (4.2 V/µm), (c) and by sine wave AC electric field (60 kHz, 6 V/µm).

As seen in Fig. 7a, the dielectric data for $N_{TBe}$ and $N_{SB}$ phases more or less agree with the model. The transition from $N_{TBe}$ and $N_{SB}$ phase is clearly indicated by a sudden drop in the mode's frequency. In the range of $N_{SB}$ phase the dependence on $E$ is linear but much steeper than predicted by the model. Further increase in the field shifts the relaxation peak into range of frequencies where ionic and relaxation processes overlap with each other. However $q$ vs. $E$, can be picked up from the texture. These experimental points lie in the $N^*_{SBp}$ phase region, where the helical structure is gradually being unwound. As seen in Fig. 7a, the relaxation rate, $f_S$, of the low mode shows much stronger dependence on $E$ than expected from the model in $N_{SB}$ (or $N^*_{SBp}$) phase. It needs to be noted that the model is given for the bulk sample and not confined to the boundaries of the cell. Texture in figure 7b shows transition from the periodic $N_{SB}$ phase to a uniform (unwound) $N_p$ structure. Figure 7c shows texture of the cell even under higher AC field of 60 kHz. Sharp striped domains indicate uniform $N_p$ structure separated out into domains of various orientations. On increasing $E$ further, the period stripe widths are increasing linearly, as already found for achiral dimer with negative dielectric anisotropy [28].

We analyse distortions in the $N_{TB}$ structure as a function of $E$ in different frequency ranges of the nematic phases of a bent core liquid crystal. The field stabilises the macroscopic heliconical structure and the average dielectric energy is minimum for orientations of the optic axis. For frequencies of ~$10^5$ Hz, the dielectric effect is dominant due to a finite time average of the torque for the quadratic effect. We observe fast relaxation mechanisms in $N_{TB}$, ranging from 2 μs to 0.7 μs under cooling from 438K to 425K. At the microsecond time scale, the field is varied rapidly such that complications in dynamics are avoided by assuming that helicoidal structure is dynamically frozen at its field-off state. The relaxation rate is found to be proportional to the elastic torque, the latter is primarily dependent on $\sin^2\theta$, calculated independently from the dielectric strength data. It is interesting to note that the $\theta$ is finite for temperatures few degrees above $N_{TB}$-N phase transition temperature coinciding with existence of the high temperature $N_{TB}$ phase. For frequencies below $10^4$ Hz, the time average of torque is significantly large due to the longer time period and larger $\Delta\varepsilon$. The electric field persists long enough for the macroscopic effects for the reorganisation of pseudo-layered structure of $N_{TB}$ observable. For compounds with negative $\Delta\varepsilon$, transformations of the initial $N_{TB}$ phase into three subphases are possible. These are the one-dimensional modulated nematic structures (ODMNS), where nematic splay bend, $N_{SB}$, exists over a wide range of electric fields [14]. Results from dielectrics for $N_{TBe}$ and $N_{SB}$ qualitatively agree with those from the model, the transition from $N_{TBe}$ and $N_{SB}$ phase is clearly observed by the drop in the wave vector, $q$, occurring at the transition. In the range of fields for $N_{SB}$, the dependence of $q$ on $E$ is linear but a lot steeper than predicted by the model. On increasing $E$, pitch of the periodic structure increases as predicted, helical structure is gradually unwound and size of the domain approaches tens of μm for $E$ > 5V/μm. Frequencies of the mode depend on $E$, the sample alignment and the surface anchoring strengths.

**Acknowledgement:**

Work of Dublin group was supported by 13/US/I2866 from the Science Foundation of Ireland as part of the US–Ireland Research and Development Partnership program jointly administered with the United States National Science Foundation under grant number NSF-DMR-1410649. The authors than Dr S. P. Sreenilayam for making preliminary measurements.

## REFERENCES

[1] V.P. Panov, M. Nagaraj, J. K. Vij, Yu. P. Panarin, A. Kohlmeier, M. G. Tamba, R. A. Lewis, and G. H. Mehl
Phys. Rev. Lett. **105**, 167801 (2010).

[2] V. Borshch, Y.-K. Kim, J. Xiang, M. Gao, A. Jakli, V. P. Panov, J. K. Vij, C. T. Imrie, M. G. Tamba, G. H. Mehl, and O. D. Lavrentovich, Nat. Commun. **4**, 2635 (2013).

[3] D Chen, J. H. Porada, J. B. Hooper, A. Klittnick, Y. Shen, , M. R. Tuchband, E. Korblova, D. Bedrov, D. M. Walba, M. A. Glaser, J. E. Maclennan and N. A. Clark, PNAS. **110**, 15931 (2013).

[4] C. Meyer, G. R. Luckhurst, and I. Dozov. Phys Rev Lett. **111**, 067801 (2013).

[5] C. Meyer and I. Dozov, Soft Matter **12**, 574 (2016).

[6] H. J. Deuling, Mol. Cryst. Liq. Cryst. **19**, 123 (1972)

[7] S. M. Shamid, S. Dhakal and J. V. Selinger, Phys Rev E **87**, 052503(2013).

[8] G. Shanker, M. Nagaraj, A. Kocot, J. K. Vij, M. Prehm and C. Tschierske, Adv. Funct. Mater. **22**, 1671 (2012).

[9] N. Vaupotic, M. Cepic, M. A. Osipov and E. Gorecka. Phys. Rev. E **89**, 030501 (2014).

[10] S. Garoff and R. B. Meyer, Phys Rev Lett. **38**, 848 (1977).

[11] J. S. Patel, R. B. Meyer, Phys Rev Lett. **58**,1538 (1987).

[12] C. Meyer, Liq. Cryst. 43, 2144 (2016).

[13] A. Matsuyama J. Phys. Soc. Jpn. **85**, 114606 (2016).

[14] G. Pająk, L. Longa and A Chrzanowska arXiv:1801.00027v1 [cond-mat.soft] 29 Dec 2017.

[15] L. Longa, G. Pająk, Phys. Rev. E **93**, 040701(R), 2016],

[16] R Balachandran, V. P. Panov, J. K. Vij, G. Shanker, C. Tschierske, K. Merkel and A. Kocot, Phys. Rev. E **90**, 032506 (2014).

[17] H. Toriyama, S. Sugimari, K. Moriya, D. A. Dunmur, and R. Hanson, J. Phys. Chem. **100**, 307 (1996).

[18] W.T. Coffey and Y. P., Kalmykov, Adv. Chem. Phys., 113, 487 (2000); W.T. Coffey Y. P. Kalmykov and J. T. Waldron, Liq. Cryst., 18, 677 (1995)

[19] V. P. Panov, R. Balachandran, M. Nagaraj, J. K. Vij, M. G. Tamba, A. Kohlmeier, and G. H. Mehl *Appl. Phys. Lett*. 99, 261903 (2011).

[20] S. P. Sreenilayam, V. P Panov, J. K. Vij and G Shanker, Liq. Cryst. 44, 244 (2017).

[21] Z Parsouzi,., S. M. Shamid, V. Borshch, P. K. Challa, A. R. Baldwin, M. G. Tamb, C. Welch, G. H. Mehl, J. T. Gleeson, A., Jakli, O. D., Lavrentovich, D.W Allender., J. V Selinger and S. Sprunt, Phys. Rev. X 6, 021041 (2016).

[22] D. Chen, M. Nakata, R. Shao, M. R. Tuchband, M. Shuai, U. Baumeister, W. Weissflog, D. M. Walba, M. A. Glaser, J. E. Maclennan, and N. A. Clark *Phys. Rev*. E **89**, 022506 (2014).

[23] D. M. Agra-Kooijman, G. Singh, M. R. Fisch, M. R. Vengatesan, J.-K. Song, and S. Kumar, Liq. Cryst., **44**, 191 (2017).

[24] C. J. Yun, M. R. Vengatesan, J. K. Vij, and J. -K Song, App. Phys. Lett. **106**, 173102 (2015).

[25] V. P. Panov J. K Vij. and. G. H. Mehl, Liq. Cryst. 44, 147 (2017)

[26] V. P. Panov M. C. M. Varney, I. I. Smalyukh, J. K. Vij, M. G. Tamba,. and G. H. Mehl, Mol. Cryst. Liq. Cryst. **611**,180 (2015).

[27] P. G. de Gennes and J. Prost, The Physics of Liquid Crystals, 2nd ed. (Clarendon Press, 1993).

[28]. V. P Panov, R. Balachandran, J. K. Vij M. G. Tamba, A. Kohlmeier, and G. H. Mehl, Appl. Phys. Lett. **101**, 234106 (2012).